\documentclass[12pt,preprint]{aastex}

\usepackage{graphicx}

\newcommand{\HI}{{\sc H\,i}}
\newcommand{\HII}{{\sc H\,ii}}
\newcommand{\kms}{\rm km\ s^{-1}}
\newcommand{\ugca}{UGCA~86}

\title{\HI\ Distribution and Kinematics of \ugca}  

\author{J. M. Stil}
\affil{Department of Physics and Astronomy, University of Calgary}
\affil{2500 University Drive NW, Calgary, AB, T2N 1N4, Canada}
\email{stil@ras.ucalgary.ca}
\author{A. D. Gray}
\affil{Dominion Radio Astrophysical Observatory}
\affil{Herzberg Institute of Astrophysics, National Research Council Canada}
\affil{P. O. Box 248, Penticton, BC, V2A~6J9, Canada}
\email{Andrew.Gray@nrc-cnrc.gc.ca}
\and
\author{J. I. Harnett}
\affil{Faculty of Engineering, University of Technology, Sydney}
\affil{P. O. Box 123, Broadway, NSW, 2007, Australia}
\email{jules@eng.uts.edu.au}

\shorttitle{\HI\ Distribution and Kinematics in \ugca}
\shortauthors{Gray, Stil, \& Harnett}

\begin{document}

\begin{abstract}
We present 21-cm \HI\ line and 408 MHz and 1.4 GHz continuum
observations of the Magellanic dwarf galaxy \ugca, made with the DRAO
synthesis telescope. \ugca\ is detected in the continuum at both
frequencies, with 408 MHz flux density $S_{408} = 120 \pm 30\ \rm
mJy$ and 1.4 GHz flux density $S_{1400} = 79 \pm 3\ \rm mJy$.  The
\HI\ structure of \ugca\ is complex, with two separate components: a
rotating disk and a highly elongated spur that is kinematically
disjunct from the disk.  The \HI\ disk is centered on the optical
galaxy with similar axial ratio and orientation of the major axis.  An
area of the disk with a peculiar velocity of $\sim 25\ \kms$ relative
to the regular rotation of the disk is found on the southern side,
where most of the star formation is concentrated.  The spur is seen
along the minor axis of \ugca\ and overlaps in part with the disk.
Towards the optical center of \ugca, the velocity difference between
the spur and the disk is $40\ \kms$, about one third of the rotation
velocity of the \HI\ disk at 6 kpc from the center. This implies a
large radial component of the orbital velocity of the spur, and
therefore a significantly non-circular orbit. The median \HI\ velocity
dispersion of the disk is $8.8\ \kms$, similar to other (dwarf)
galaxies.  The \HI\ velocity dispersion of the spur varies from $10\
\kms$ to $30\ \kms$. A possible tidal origin of the spur is considered
in view of the proximity of the large Scd galaxy IC~342. However, the
orientation of the spur along the minor axis and its spatial overlap
with the disk suggest that the spur is located far outside the plane
of the \HI\ disk. No evidence is found that the outer \HI\ disk is
warped, which poses a problem for the interpretation of the spur as a
tidal tail induced by IC~342. Detailed modelling of the IC~342/\ugca\
system will be required before a tidal origin of the spur can be
dismissed conclusively. The possibility that the spur is part of the
nascent cloud of \ugca\ or the remains of a small dwarf galaxy is
presented as an alternative interpretation.
\end{abstract}

\keywords{galaxies: dwarf -- galaxies: kinematics and dynamics --
galaxies: structure -- galaxies, individual: UGCA~86, IC~342 -- radio
lines: galaxies}

\maketitle

\section{Introduction}

\ugca\ is a Magellanic spiral (Sm) galaxy originally catalogued from
optical plates by \citet{Nilson74}.  It is also known as A0355+66
\citep{KKT79}, and was discovered independently through its \HI\
emission by Rots (1979), who referred to it as A0355.  \ugca\ was
proposed as a member of the Local Group based on an analysis of a
colour-magnitude diagram \citep{SH91}, but photometry of the brightest
stars place it at a distance of 2.6\,Mpc, in the IC~342/Maffei Group
\citep{KT93,Kar97}. \ugca\ is located $94'$ southeast of the center of
the Scd galaxy IC 342, implying a minimum separation of 71 kpc for the
assumed distance of $2.6\ \rm Mpc$. The adopted distance is somewhat less
than the 3.3 Mpc distance of IC~342 \citep{Saha2002}, but in view of
the uncertainties, \ugca\ is likely a satellite of IC 342. Its present
distance from IC 342 is at least 50\% larger than the distance between
the Milky Way and the Magellanic clouds.

Optical studies are complicated by the low Galactic latitude of this
galaxy ($b\simeq+11^\circ$). \citet{HM95} found reddening $E(B-V) =
0.59 \pm 0.23$ for \HII\ regions in \ugca, with no evidence for
significant extinction inside \ugca. Surface photometry in $V$ and $I$
by \citet{BM99} shows a nearly exponential surface brightness profile
with central surface brightness $\mu_{\rm 0,V} = 20.14\ \rm mag\
arcsec^{-2}$ and integrated magnitudes $V_{\rm T}=11.89 \pm 0.07$,
$(V-I)_T = 1.49 \pm 0.1$. Applying a standard Galactic extinction
curve, the absolute magnitude in the $V$ band corrected for extinction
is $M_V = -17.0$ with an uncertainty of at least one magnitude. The
extinction-corrected central surface brightness is $\mu_{\rm 0,V} =
18.3\ \rm mag\ arcsec^{-2}$.  The shape of the optical isophotes is
elliptical with mean axial ratio $b/a = 0.80 \pm 0.02$ and major axis
in P.A. $26\fdg8 \pm 1\fdg8$. The optical axial ratio implies an inclination
$i = 38^\circ$ for an intrinsic axial ratio $(b/a)_0 = 0.2$.

\citet{HM95} detected numerous \HII\ regions in \ugca\ adding up to a
total star formation rate of $0.014\ \rm M_\odot\ yr^{-1}$.  Most of
the star formation is concentrated in three \HII\ region complexes
southeast and southwest of the center of \ugca.  The most prominent
star formation complex is the ``cometary'' component to the south east
(see Figure~20 of Buta \& McCall 1999).  That component is catalogued
as VIIZw~009, and is well resolved into stars.  It was suggested to be
a separate, disturbed galaxy by \citet{SH91}, but has since been shown
to be a part of \ugca\ \citep{Ri91,KT93,HM95}. \citet{BM99} likened
the cometary component to the 30~Doradus complex in the Large
Magellanic Cloud.  The metallicity of \HII\ regions 81 and 101 in
\citet{HM95} is ${\rm log}(O/H) = -4.2$, comparable to the Large Magellanic
Cloud \citep{Kurt98}. Both these \HII\ regions are associated with the
cometary knot southeast of the center of \ugca.  \citet{Richter98}
argued that the cometary component is a more recent burst of star
formation than the central part of the galaxy, and speculated that it
was triggered by infall of gas. Infrared work on this galaxy has
concentrated on the prominent \HII\ regions in the south-west regions
of the galaxy, with similar conclusions \citep{Braun00}.  An X-ray
source is located near the core (ROSAT source 1RXP J035951+6708.6),
and has been suggested as being a low-mass X-ray binary
\citep{Richter98}.

\citet{Rots79} showed that \ugca\ is rotating, with a perturbed
velocity field, and identified it as a Magellanic dwarf irregular
galaxy.  His \HI\ mass reduced to a distance of 2.6 Mpc is $1.2\
\times\ 10^9\ \rm M_\odot$.  \citet{Rots79} reported a systemic
velocity $v_{\rm sys}= 85\ \pm\ 5\ \kms$.  Other authors found a
significantly smaller value $v_{\rm sys}=72\ \kms$ \citep{Fisher81}
and $v_{\rm sys} = 67 \pm 4\ \kms$ \citep{Bottinelli90}. The Galactic
coordinates of \ugca\ are $(l,b) = (139\fdg76, +10\fdg65)$. At this
longitude, positive velocities are forbidden by Galactic
rotation. Confusion with foreground Galactic \HI\ is not a problem at
velocities exceeding $10\ \kms$.  The systemic velocity of \ugca\
close to the systemic velocity of IC 342 ($33\ \pm\ 4\ \kms$)
\citep{Bottinelli90}, which adds to the suspicion that \ugca\ is a
satellite of IC 342.  \citet{Rots79} further suggested that \ugca\ is
tidally interacting with IC~342 as a means of explaining the
perturbations in the velocity fields of both galaxies.  A suggestion
by \citet{Newton80} that UGC~2826 might be the perturber for IC~342
has since been ruled out by the high radial velocity of that galaxy
\citep{Strauss92}.

Rots' low-resolution ($10'$, $22\ \kms$) \HI\ results are the only
published \HI\ maps of \ugca\ to date.  To address this lack and
investigate the possible link between \ugca\ and IC~342, we have
mapped the decimetre radio continuum and neutral hydrogen distribution
of both galaxies with $1'$ spatial and $5.2\ \kms$ velocity
resolution, using the Synthesis Telescope at the Dominion Radio
Astrophysical Observatory (DRAO).

\section{Observations and Reduction}

Data from a pointing centered on the optical disk of \ugca\  were
acquired simultaneously in the \HI-line and in continuum bands at
1420 MHz and 408 MHz (wavelengths of 21 and 74\,cm, respectively).
Data were also collected for IC~342 in a separate but overlapping
pointing (the 1420\,MHz primary beam diameter of the DRAO array is
$107'$ to half power; at 408\,MHz it is $332'$). Analysis of the 
IC~342 field is deferred to another paper.

Twelve sets of spacings were used, giving baselines from 12.9\,m to
604.3\,m at 4.3\,m intervals, and yielding sensitivity to spatial
structures from $1'$ to $\sim1^\circ$ at 1.4\,GHz (and about 3.5 times
those values at 408\,MHz).  These data were collected during 2000
October--December.  Some brief technical details of the observations
and reduction are given in the following sections; refer to Landecker
et~al.\ (2000) for full details of the Synthesis Telescope.  All
images presented below have been corrected for primary beam
attenuation.

\subsection{Neutral-Hydrogen Spectral Data}

\HI\ line data were collected in 256 contiguous channels over a 4\,MHz
band, spanning 844\,km\,s$^{-1}$ (3.3\,km\,s$^{-1}$ per channel) with a
velocity resolution of 5.2\,km\,s$^{-1}$.  The observation was centered
at $+31$\,km\,s$^{-1}$, the heliocentric velocity of IC~342; \ugca\  is
nominally at +67\,km\,s$^{-1}$, which is well within the same band.
Both right- and left-hand circular polarizations were recorded, with the
two being averaged to make the final images, resulting in an rms noise
per channel of 9.5\,mJy\,beam$^{-1}$.  To remove continuum sources,
channels free from hydrogen emission and band-edge effects were selected
from each end of data-cube and averaged together (120 such channels were
used), then the result was subtracted from each channel image.

The spatial resolution achieved from uniformly-weighted visibilities
was $49'' \times 54''$ at P.A. $-0^\circ\!\!.3$; the data presented in
this paper have had a radial taper applied to increase the brightness
temperature sensitivity, and have a beam of $59'' \times 65''$ at P.A.
$-0\fdg6$.  A spectral line cube tapered to $1\farcm4$ resolution was
made to obtain a better signal to noise ratio for faint extended
emission.  The well-sampled aperture plane and inherent low
signal-to-noise meant that it was not necessary to deconvolve the
channel images (a standard practice at DRAO).  Missing short spacings
cause a mild negative bowl.

\subsection{Continuum Data}

The 1.4\,GHz continuum system records full polarization data in four
7.5\,MHz bands, pairs of which flank the spectrometer band.  Data from
the four bands were gridded separately onto a common aperture plane to
make a single image in each of the four Stokes parameters.  The
synthesized beam at 1.4\,GHz was the same as for the radially tapered
\HI\ data described above, and the rms noise was
0.25\,mJy\,beam$^{-1}$.  No significant linear or circular
polarization was seen from \ugca.

The 408\,MHz continuum system records data over a 3.5\,MHz bandwidth, in
right-hand circular polarization only.  In the absence of circularly
polarized signal, which is generally the case, this may be interpreted
as Stokes~$I$.  At 408\,MHz the tapered synthesized beam was $3'\!\!.3
\times 3'\!\!.8$ at P.A. $-3^\circ\!\!.5$, and the confusion-limited rms
noise achieved was 4\,mJy\,beam$^{-1}$.

The data were processed in DRAO's \texttt{MADR} package.  The raw
Stokes~$I$ image at each frequency was deconvolved using {\sc CLEAN},
and self-calibrated to remove artefacts arising from variations of
amplitude and phase on short time-scales during the observation.

\section{Results and Analysis}

\subsection{Neutral-Hydrogen Spectral-Line}

\begin{figure}
\caption{ \HI\ channel maps of \ugca\ at $1\farcm4$
resolution. Grayscales are linear from 0 mJy/beam (white) to 400
mJy/beam (black). Contours are drawn at $-3$, $3$, $5$, $7$, $11$, $19$ and
$35$ times the r.m.s. noise per channel ($9.5\ \rm
mJy/beam$). Velocities in $\kms$ are shown in the upper left corner of
each panel.  Channels near zero velocity are affected by Galactic
foreground emission. The size of the synthesized beam is shown in the
bottom right panel.  }
\label{chanmaps-fig}
\end{figure}
%

Figure~\ref{chanmaps-fig} shows the \HI\ channel maps of \ugca\ at
positive velocities. Emission of \ugca\ is blended with Galactic
emission at velocities below $\sim\!+10\ \kms$. The location of the brightest
emission associated with \ugca\ changes gradually with velocity in a
way consistent with a rotating disk. At velocities beyond $50\ \kms$
the structure in the channel maps is more complex.  In addition to the
brighter component, a faint spur of emission extends towards the
northwest.

\begin{figure}
\caption{Contours of the \HI\ column density for \ugca\  superimposed
  on a grayscale of the same data.  The contours are at
  $2\times10^{20}$\,cm$^{-2}$ intervals starting at
  $2\times10^{20}$\,cm$^{-2}$.  The grayscale is linear from 0 to
  $2.5\times10^{21}$\,cm$^{-2}$.}
\label{gray:hicolumn}
\end{figure}
\begin{figure}
\caption{
  Contours of the velocity field for \ugca\  overlaid on a
  grayscale image of the same data.  The contours are at $15\ \kms$
  intervals starting at $-10\ \kms$.  Selected contours are
  labeled with the corresponding velocity in $\kms$.}
\label{gray:velfield}
\end{figure}

Figures~\ref{gray:hicolumn} and \ref{gray:velfield} show the \HI\
column density and velocity field, respectively, for \ugca.  These
images were constructed from a cube in which the emission from \ugca\
was manually identified in each channel image and selectively masked
to exclude areas with no signal.  In this way we minimized the adverse
effects of noise in empty regions, and the contaminating Galactic \HI\
emission at velocities below $\sim+10\ \kms$. In these contaminated
channels we also examined position-velocity slices though the
data-cube to aid in the identification of emission from \ugca, which
has higher velocity dispersion than the Galactic emission.  The total
\HI\ mass of \ugca\ derived from this masked dataset is $0.66\times
10^9\ \rm M_\odot$.  This is significantly smaller than the value
found by \citet{Rots79}.  This difference cannot be attributed to
resolution of the extended emission in Rots' map by the shortest
baselines of the DRAO interferometer. The present data lack the
sensitivity to detect the extended low column density emission in
Rots' map, and the modest negative bowl has a negative effect on the
flux when integrating over a large area.  Differences in the treatment
of confusing Galactic emission may also have contributed. Note that
the extended emission in the first panel of Figure~\ref{chanmaps-fig}
is considered Galactic here.

The distribution of neutral hydrogen has a maximum column density of
$2.5\times10^{21}$\,cm$^{-2}$.  The distribution is approximately
elliptical in outline for column densities higher than $6 \times
10^{20}$\,cm$^{-2}$ (Figure~\ref{gray:hicolumn}), with an extent of
$13\farcm4 \times 9\farcm6$ at the $5 \times 10^{20}\ \rm cm^{-2}$
level, oriented at P.A.  $50^\circ$. The axial ratio of \HI\ column
density contours in the range $4 \times 10^{20}\ \rm cm^{-2}$ to $8
\times 10^{20}\ \rm cm^{-2}$, excluding the spur, is $(b/a)_{\rm HI} =
0.72 \pm 0.02$. This part of \ugca\ is centered on the optical galaxy
with similar axial ratio and orientation of the apparent major axis
\citep{Rots79,BM99}. The velocity field in this area shows a
well-defined rotational velocity gradient with the receding side of
the galaxy in P.A. $287^\circ$. Both the hydrogen and optical disks
are centered at ($\alpha_{\rm J2000}$, $\delta_{\rm J2000}$) =
($03^{\rm h}\,59^{\rm m}\,50^{\rm s}\!\!.5$, $+67^\circ\,08'\,37''$).
There are three prominent ``holes'' in the \HI\ at ($03^{\rm
h}\,59^{\rm m}\,54^{\rm s}\!\!.5$, $+67^\circ\,07'\,37''$), ($04^{\rm
h}\,00^{\rm m}\,04^{\rm s}\!\!.5$, $+67^\circ\,08'\,37''$), and
($04^{\rm h}\,00^{\rm m}\,28^{\rm s}\!\!.5$, $+67^\circ\,11'\,57''$),
in which the column density drops below $4\times10^{20}$\,cm$^{-2}$.

At low column densities the narrow \HI\ spur noticed in
Figure~\ref{chanmaps-fig} extends towards the northwest (the
approximate direction of IC 342), up to $12'$ (9 kpc) from the center
of \ugca.  This structure is also visible in the low-resolution column
density map of \citet{Rots79}. A faint extension of the \HI\ isophotes
is seen on the opposite side of the disk near 04$^{\rm h}$~01$^{\rm
m}$, $+67^\circ$~$04'$. The suggestion of an elongated structure that
partially overlaps with the \HI\ disk on the sky will be confirmed
later by position-velocity diagrams and \HI\ line profiles.  The
kinematics of the spur are ordered, with lower velocities occurring on
the eastern side of the spur. Note that the velocity changes more
rapidly along the minor axis of the spur than in the disk, and that
the (apparent) velocity gradient has a different direction.

\subsubsection{Rotation curve of the disk}

The velocity field (Figure~\ref{gray:velfield}) clearly shows the
regular rotation of the disk. However, the spur affects the velocity
field as it overlaps in position with the disk. The approaching side
of the disk is also affected by confusion with Galactic \HI. A tilted
ring fit \citep{warner1973} to the velocity field of
Figure~\ref{gray:velfield} was attempted to obtain the rotation curve
of the disk. However, the results of the fit to the velocity field are
skewed towards smaller rotation velocities because of the
contaminating emission of the spur. This will be illustrated further
in Section~\ref{spur-section}.

The best representation of the rotation curve of the disk was obtained
by fitting the shape of the position-velocity diagram though the
kinematic major axis of the disk. The P.A. of the kinematic major axis
and an initial rotation curve were determined from the tilted ring
fits.  The rotation velocities of the annuli and the systemic velocity
were adjusted manually to force the rotation curve to follow the
bright curved ridge of emission in Figure~\ref{gray:rotcurve}. The
resulting rotation curve is listed in Table~\ref{rotcur-table}. The
total mass of \ugca\ within the outermost radius of the rotation curve
is $2.1 \times 10^{10}\ \rm M_\odot$.

The inclination of the \HI\ disk is uncertain. The adopted inclination
$i_{\rm HI} = 45^\circ$ is based on the \HI\ axial ratio $(b/a)_{\rm
HI} = 0.72$ and an intrinsic axial ratio $(b/a)_0 = 0.2$. If the
optical inclination $i_{\rm opt} = 38^\circ$ is adopted, the
velocities are 15\% higher than listed in
Table~\ref{rotcur-table}. The systemic velocity found here is
consistent with the values of \citet{Fisher81} and
\citet{Bottinelli90}, but significantly smaller than the $85\ \kms$
found by \cite{Rots79}. This systematic difference may be the result
of Rots discarding channels affected by Galactic emission, and
inclusion of the spur.

A significant misalignment between the P.A. of the apparent major axis
and the kinematic major axis is observed in \ugca. A small but
significant misalignment between the kinematic and morphological major
axis is not exceptional in dwarf galaxies.  \citet{Skillman88}
interpreted a similar misalignment in the dwarf irregular galaxy Sex~A
as evidence that the gas in the disk describes non-circular orbits.

Figure~\ref{holeslice-fig} shows a position-velocity slice in
P.A. $45^\circ$, approximately along the apparent major axis of the
\HI\ disk.  A small area $3\farcm5$ southwest of the center of \ugca\
displays a significant departure from the regular rotation pattern of
the disk. Locally the line of sight velocity deviates by $25\ \kms$
from the value expected from the rotation curve model of
Table~\ref{rotcur-table}. The area where the velocity deviates more
than $20\ \kms$ from the rotation model is indicated by the ellipse in
the upper panel of Figure~\ref{holeslice-fig}. The \HI\ mass of this
area integrated over the velocity range $40\ \kms$ to $90\ \kms$ is
$2.9 \times 10^7\ \rm M_\odot$. Accounting for 30\% helium by mass,
the kinetic energy associated with this region is $2.4 \times 10^{53}\
\rm ergs$. This energy is equivalent with the mechanical energy
released by 240 supernovae.

\begin{deluxetable}{ccccc}
\tablecolumns{5}
\tablewidth{0pc} 
\tablecaption{Rotation curve of \ugca}
\tablehead{
\colhead{radius} & \colhead{Velocity} & \colhead{inclination} & \colhead{P.A.} & \colhead{$v_{\rm sys}$}\\
\colhead{$''$} & \colhead{$\kms$} & \colhead{$^\circ$} & \colhead{$^\circ$} & \colhead{$\kms$}\\
}
\startdata
 \phn60   & \phn35  $\pm$ 5 & 45 & 287 $\pm$ 2 & 72 $\pm$ 5\\ 
    120   & \phn65  $\pm$ 5 & 45 & 287 $\pm$ 2 & 72 $\pm$ 5\\ 
    180   & \phn87  $\pm$ 5 & 45 & 287 $\pm$ 2 & 72 $\pm$ 5\\ 
    240   &  103  $\pm$ 5   & 45 & 287 $\pm$ 2 & 72 $\pm$ 5\\ 
    300   &  114  $\pm$ 5   & 45 & 287 $\pm$ 2 & 72 $\pm$ 5\\ 
    360   &  119  $\pm$ 5   & 45 & 287 $\pm$ 2 & 72 $\pm$ 5\\ 
    420   &  122  $\pm$ 5   & 45 & 287 $\pm$ 2 & 72 $\pm$ 5\\ 
    480   &  122  $\pm$ 5   & 45 & 287 $\pm$ 2 & 72 $\pm$ 5\\ 
\enddata
\label{rotcur-table}
\end{deluxetable}

\begin{figure}
\caption{Rotation curve of \ugca, superimposed on a
  position-velocity slice along the dynamical major axis (position
  angle $287^\circ$). Position offset increases in the direction of
  right ascension. Grayscales are linear from 0 to 150
  mJy/beam. Circles indicate velocities of the rotation curve listed
  in Table~\ref{rotcur-table}. Emission of the spur is also visible at
  velocities between 0 and $\sim$ +75 km\,s$^{-1}$.}
\label{gray:rotcurve}
\end{figure}

\begin{figure}
\caption{ Position-velocity slice along the apparent major axis of
  \ugca.  The top panel shows the location of the slice on the \HI\
  column density map.  The ellipse indicates an area in the disk where
  the velocity deviates more than $20\ \kms$ from the line of sight
  velocity expected from the rotation curve and the inclination of the
  disk.  The bottom panel shows the position-velocity slice. The
  curved line indicates the velocities calculated from the rotation
  curve in Table~\ref{rotcur-table}. Grayscales in the bottom panel
  are linear from 0 to 150 mJy/beam.  }
\label{holeslice-fig}
\end{figure}

\subsubsection{Kinematics and mass of the spur}

\label{spur-section}

Position-velocity slices parallel to the major axis (P.A.
322$^\circ$) and the minor axis (P.A. 232$^\circ$) of the spur are
shown in Figure~\ref{gray:majorslice} and Figure~\ref{gray:minorslice}
respectively. These slices were taken from the $1\farcm4$ resolution
cube to show faint emission in the spur.  Galactic foreground \HI\ is
seen at negative velocities and close to zero. 

The spur is clearly visible in the velocity range $+50\ \kms$ to
$+140\ \kms$, far outside the velocity range permitted by Galactic
rotation assuming circular orbits and a flat rotation curve. It is
possible in principle that the spur is \HI\ in the foreground that
does not follow Galactic rotation. We note that the difference between
heliocentric velocity and velocity with respect to the Local Standard
of Rest (LSR) is only $2\ \kms$ for \ugca.

One possibility is that the spur is a high-velocity cloud (HVC) along
the line of sight. The HVC complexes nearest to the line of sight are
complex A and complex H in the velocity range $-80\ \kms > v_{\rm LSR}
> -210\ \kms$ \citep{Wakker1991}. Association of the spur with these
complexes is dismissed because of the large velocity
difference. Instead, the spur could be considered a compact
high-velocity cloud (CHVC), not associated with a large
complex. However, the all-sky distribution of velocities of CHVCs in
the LSR reference frame is dominated by the orbital velocity of the
LSR around the Galactic center. The velocity of the LSR has a large
component in the direction of \ugca, and no CHVCs with positive
velocities in the LSR frame have been found in this part of the sky
\citep[and references therein]{deheij2002}. The probability that an
extraordinary CHVC with a positive velocity is found in this location
is very small. For this reason, the interpretation of the spur as a
CHVC is rejected.

\citet{lockman2002} reported a population of \HI\ clouds in the
Galactic halo with velocities up to $50\ \kms$ beyond that allowed by
Galactic rotation. Some of these clouds would be detectable in the
present data \citep{lockman2003}.  The spur is observed at much higher
forbidden velocities. Also, the velocity gradient of the spur and the
width of the \HI\ line profile in the spur do not match the narrow
line widths observed in these halo clouds
\citep{lockman2002,lockman2004}.

A distinct velocity gradient is observed along both the major and the
minor axes of the spur. This indicates that the actual velocity
gradient is not aligned with either the major axis or the minor
axis. The direction of the actual velocity gradient across the spur
was determined by interactively changing the P.A. of position-velocity
slices until no velocity gradient was discernible. This gives the
P.A. of the direction perpendicular to the velocity gradient. The
velocity gradient across the spur was found to be $\nabla v_{\rm spur} =
20.4\ \kms\ arcmin^{-1}$ in P.A. $75^\circ$.  Evidence for a reversal
of the velocity gradient is seen in slice D of
Figure~\ref{gray:majorslice} towards the extreme end of the spur.

Figure~\ref{gray:majorslice} also shows that the spur does not begin
at the northwest edge of the disk, but rather overlaps on the sky with
the \HI\ disk. The \HI\ line profiles are double-peaked or blended
where the spur and the disk overlap as illustrated in
Figure~\ref{centerprofiles}. Double-peaked \HI\ line profiles with a
velocity separation of $\Delta v = 40\ \kms$ are seen near the center
of \ugca. Southeast of the center of \ugca\  the line profiles are
blended so much that a kinematic separation is not possible. The
asymmetric line profiles on the southeast side do suggest that the
spur extends across the disk. On the extreme east side of \ugca,
the disk itself is difficult to disentangle from Galactic \HI.

The difference $\Delta v$ in the line of sight velocity of the spur
and the disk towards the center of \ugca\  is a direct measurement of
the velocity component of spur gas in the direction of the center of
mass of \ugca, as defined by the optical galaxy. The radial
component of the spur gas in this direction is a significant fraction,
approximately one third, of the circular velocity at the last measured
point of the rotation curve (Table~\ref{rotcur-table}). As the
rotation curve appears to turn over around a radius of $300''$, the
velocity of a circular orbit at larger radii is not expected to exceed
$122\ \kms$ by a significant amount.  Therefore, the radial component
of the velocity of this part of the spur is at least one third of the
local circular velocity, independent of the actual distance from the
center of \ugca. This significant radial component of the orbital
velocity is a direct indication that the orbits of gas in the spur are
significantly non-circular with respect to the center of \ugca.

A determination of the mass of the spur requires decomposition of the
line profiles in a disk and a spur component in places where the spur
and the disk overlap on the sky.  Figure~\ref{centerprofiles} shows
that a decomposition of the line profiles into Gaussians will not
provide a unique solution in many places. This ambiguity was addressed
by subtracting a sequence of models for the \HI\ disk from the data.
These models were generated in three steps.

In the first step, a decomposition into at most two Gaussians was
attempted at every location. The strongest Gaussian component was
tentatively assigned to the disk.  If no clear decomposition could be
made, the profile was skipped in this step.  In the second step, the
central velocity of the Gaussian assigned to the disk was compared
with the line of sight velocity expected from the rotation curve, the
inclination of the disk, and the location in the disk. If the velocity
difference exceeded a certain threshold, the fit was
rejected. Thresholds of 0, 5, 10, 20 and 50 $\kms$ were applied to
generate a sequence of disk models that assign less weight to the
rotation curve and increasing weight to the Gaussian decomposition.
In the third step, rejected profiles and profiles that could not be
decomposed into separate components in step 1, were replaced by a
model Gaussian. This model Gaussian has a central velocity calculated
from the rotation curve, amplitude equal to the observed intensity at
the central velocity, and dispersion equal to the median velocity
dispersion of the \HI\ disk ($13\ \kms$). Each model was subtracted
from the data. The flux of the residual emssion was measured to obtain
the mass of the spur. At this time, the anomalous velocity region
shown in Figure~\ref{holeslice-fig} was masked out. It is not included
in the mass of the spur.  The \HI\ mass of the spur was found to be in
the range $4.8\ \times\ 10^7\ \rm M_\odot$ to $1.2\ \times\ 10^8\ \rm 
M_\odot$.  The \HI\ mass of the spur is approximately 10\% of the \HI\
mass of \ugca, but well within the range of \HI\ masses of nearby dwarf
galaxies \citep{karachentsev2004}.

\begin{figure}
\caption[]{Position velocity slices along the major axis of the spur
  (P.A. 322$^\circ$) extracted from the $1\farcm4$ resolution cube. The
  location of the slices is shown on the \HI\ column density map of
  Figure~\ref{gray:hicolumn}.  Slices A and D are labeled. The short
  line perpendicular to the slices connects the locations of the zero
  points in the position-velocity diagrams.  This line corresponds
  with slice C in Figure~\ref{gray:minorslice}. Position offsets
  increase in the direction of Right Ascension. Grayscales in the
  position-velocity diagrams are linear from 0 (white) to
  75\,mJy\,beam$^{-1}$ (black).  Contours are drawn at 50, 100, 150,
  \ldots\ mJy/beam.
}
\label{gray:majorslice}
\end{figure}
\begin{figure}
\caption[]{Position velocity slices along the minor axis of the spur
  (P.A. 232$^\circ$) extracted from the $1\farcm4$ resolution
  cube. The location of the slices is shown on the \HI\ column density
  map of Figure~\ref{gray:hicolumn}.  Slices A and F are labeled. The
  short line perpendicular to the slices connects the locations of the
  zero points in the position-velocity diagrams.  This line
  corresponds with slice D in Figure~\ref{gray:majorslice}. Position
  offsets increase in the direction of Right Ascension.  Grayscales
  are linear from 0 (white) to 75\ mJy/beam (black).  Contours are
  drawn at 50, 100, 150, \ldots\ mJy/beam. The brightest emission in
  slice F is a part of the \HI\ disk.
}
\label{gray:minorslice}
\end{figure}

\begin{figure}
\caption{ \HI\ line profiles extracted from the $1'$ resolution \HI\
cube in the area where the spur and the disk overlap. Positions of the
profiles are shown as $+$ on the velocity field of
Figure~\ref{gray:velfield}.  The line profiles in the central area
show two components separated in velocity by approximately $40\
\kms$. The least redshifted component is usually associated with the
spur (see also Figure~\ref{gray:majorslice}).  }
\label{centerprofiles}
\end{figure}

\subsubsection{\HI\ velocity dispersion}

The velocity dispersion of \HI\ in the spur and the disk was
determined by fitting a single Gaussian to the \HI\ line profiles.
The $1\farcm4$ resolution data cube was regridded to 1\farcm5 pixels
to obtain a single spectrum per beam.  Line profiles with multiple
components in places where the disk and the spur overlap were excluded
from this analysis.  Measurements of the velocity dispersion in the
spur include areas west of $\alpha_{\rm J2000} = \rm 3^h 59^m$ and
north of $\delta_{\rm J2000} = 67^\circ 16'$, which are separated
spatially from the disk.  Only 12 spatially independent profiles with
sufficient signal were available in the spur. Some of the profiles in
the spur appear to be the superposition of two or more components, but
the limited signal to noise ratio does not warrant a decomposition
into multiple Gaussians.

\begin{figure}
\caption{Distribution of the \HI\ line-of-sight velocity dispersion in the
  disk of \ugca\  (top) and in the \HI\ spur (bottom).}
\label{gray:histo}
\end{figure}

Figure~\ref{gray:histo} shows histograms of the \HI\ velocity
dispersion in the disk and in the spur.  The median error for
velocity dispersions is $1.0\ \kms$ in the disk, and
$2.2\ \kms$ in the spur.  The larger errors for the velocity
dispersion in the spur are because of the generally lower brightness
and the larger velocity dispersion.

The distribution of velocity dispersions in the disk is similar to
that of other dwarf galaxies.  The median velocity dispersion in the
disk is $13.0\ \kms$, before corrections for spectral resolution and
broadening by the rotational velocity gradient.  Even though the
number of measurements in the spur is small, Figure~\ref{gray:histo}
indicates that the distributions of velocity dispersion differ between
the spur and the disk.  The median velocity dispersion in the spur is
$19.4\ \kms$.  \citet{hunter99} found similar velocity dispersions in
the outer cloud complexes of NGC~4449.

The line of sight velocity dispersions shown in
Figure~\ref{gray:histo} have not been corrected for spectral
resolution or for a contribution of the rotational velocity
gradient. The correction for rotational broadening of the line
profiles was done following \citet{Stil2002b}.  The velocity gradient
due to rotation of the disk derived from the rotation curve is $\nabla
v_{\rm disk} = 13.6\ \kms\ arcmin^{-1}$.  The instrumental spectral
resolution is $5.2\ \kms$.  The median velocity dispersion $\sigma =
13.0\ \kms$ then results in a corrected line of sight velocity
dispersion and spectral resolution $\sigma_{\rm HI} = 8.8\ \kms$.

The velocity gradient across the spur contributes to the observed line
of sight velocity dispersion in a similar way.  For the velocity
gradient $\nabla v_{\rm spur} = 20.4\ \kms\ arcmin^{-1}$, observed
velocity dispersions of $15\ \kms$, $20\ \kms$, and $30\ \kms$ imply
intrinsic velocity dispersions of 7, 15, and 27 $\kms$
respectively. The velocity gradient over the beam contributes to the
high velocity dispersion observed in the spur, but its effect is too
small to explain the difference in the distributions of the velocity
dispersion in Figure~\ref{gray:histo}.  Adopting a single velocity
gradient for the spur is a simplification.  The smallest velocity
dispersion measured in the spur is smaller than the broadening implied
by the mean gradient.  A more detailed analysis requires observations
with higher spatial resolution.

\subsection{Continuum emission}

The 1.4 GHz radio continuum emission from \ugca\
(Figure~\ref{gray:continuum}) shows a lop-sided distribution, with
several emission peaks to the south and west of the optical center,
embedded in an extended component that covers the entire optical disk.
The strongest 1.4\,GHz continuum emission coincides with the optical
cometary component, and has a similar morphology.  Emission peaks to
the west and south-west of the optical center coincide with optical
\HII\ region complexes \citep{HM95,KM98} that are also seen in
reprocessed IRAS data \citep{Braun00}. The \HII\ region complexes
appear on the periphery of depressions in \HI\ column density.  Two
unresolved sources to the east have no optical or infrared
counterparts, and are presumably background sources. The northern one
is catalogued as NVSS~J040022+670830 \citep{Condon1998}.  No radio
counterpart to the ROSAT X-ray source 1RXP J035951+6708.6 is evident.

\ugca\  is also detected at 408 MHz. Three independent complete
synthesis observations with \ugca\  near the pointing center were
analysed. One was centered on \ugca, one on IC~342, and one
unrelated observation was available in the DRAO archive.  \ugca\ 
was detected in each of these 408 MHz observations.  The 408 MHz image
in Figure~\ref{gray:continuum} is the linear mosaic of these three
observations.

The total flux density at 1.4\,GHz is $S_{1.4} = 79\pm3$\,mJy, and at
408\,MHz is $S_{0.408} = 120\pm30$\,mJy\,beam$^{-1}$.  The implied
spectral index of \ugca\  between 408 MHz and 1.4 GHz is
$\alpha=-0.34^{+0.27}_{-0.20}$ ($S_\nu\propto\nu^\alpha$).  While this
is consistent with a nearly flat spectrum, it is likely that not all
of the emission is thermal.  The association of the compact emission
with \HII\ regions suggests that the the extended emission seen in the
21-cm continuum image may be significantly non-thermal. This is
reminiscent of the extended synchrotron halo of the post-starburst
dwarf galaxy NGC~1569, which has the same global spectral index
$\alpha = -0.36$ \citep{Israel1988}.  Extended synchrotron haloes are
a rare phenomenon in dwarf galaxies. \ugca\  is a prime candidate
dwarf galaxy with a synchrotron halo.

\begin{figure}
\caption{Contours of the 1.4 GHz continuum emission from \ugca\ 
  superimposed on (left) the \HI\ column density grayscale (linear
  from 0 to $2.5\times 10^{21}$\,cm$^{-2}$), and (right) the 408 MHz
  continuum grayscale (linear from -6 to 60 mJy/beam) with a wider
  field of view.  The black solid contours start at 0.5 mJy/beam
  (2$\sigma$), in steps of 0.75 mJy/beam.  The dashed contour is at
  $-0.5$ mJy/beam.  The white contours on the right indicate the 0.25
  and 3 mJy/beam level in the 1.4 GHz continuum image.  The hatched
  ellipses indicate the beam size of the 1.4 GHz continuum map (left)
  and the 408 MHz map (right).}
\label{gray:continuum}
\end{figure}

\section{Discussion}

\ugca\  is a member of a heterogeneous set of dwarf galaxies with
very significant departures from ordered rotation in their extended
\HI\ envelopes. These galaxies are of interest for our understanding
of the formation and evolution of galaxies because of the origin of
the chaotic kinematics. We distinguish three broad categories
that arguably share some common characteristics: 1. tidal interaction
with another galaxy; 2. collision with or capture of an intergalactic
\HI\ cloud; 3. assembly of the \HI\ disk on timescales longer than a
Hubble time (``continuing galaxy formation'' \citep{hunter98}).

In case of tidal interaction, the \HI\ displaying the peculiar
kinematics is in fact the disturbed outer \HI\ disk. The most extreme
case of tidal interaction is a merger with another galaxy. In the case
of a collision or capture of an intergalactic gas cloud or a smaller
dwarf galaxy, the gravitational force on \ugca\ may be negligible, but
the infalling material may create large holes in the gas distribution
and compress the interstellar medium to high densities in some places,
triggering star formation \citep{tenorio1986,tenorio1987}.  The case
of ongoing galaxy formation assumes that the \HI\ was always bound to
the dwarf galaxy, but has not yet been assimilated into the disk. An
interesting question is why most dwarf galaxies do appear to have
formed an \HI\ disk, or even converted all of the available gas into
stars. The main difference between a collision and ongoing
assimilation is the time scale of the process. Both may be
characterized as infall.

The relative importance of these processes is not well known.
The number of (dwarf) galaxies with chaotic kinematics in their outer
parts and the details of their kinematics provide important clues
regarding the relative importance of these processes. This may in turn
provide insight in similar processes occurring in galaxies at high
redshift.  

Although the orbits of the gas are not uniquely constrained by the
observations, some constraints may be given if a geometry is
specified. An important observation is the steep velocity gradient
along the minor axis of the spur observed in
Figure~\ref{gray:minorslice}. This velocity gradient is difficult to
explain if the the spur is a filament, with its dimension along the
line of sight equal to its minor axis on the sky. It would require a
special short-lived geometry, or a mass concentration in the spur to
balance the velocity shear and maintain its filamentary structure.  It
is more likely that the velocity gradient across the spur displays the
line of sight component of the orbital velocity of gas in a structure
that is extended along the line of sight. This argument is equally
valid whether the orbits of the spur gas are open or closed.  The
large range in the \HI\ velocity dispersion of the spur
(Figure~\ref{gray:histo}) is also most easily explained in terms of a
superposition of multiple velocity components along the line of sight.
Much of the following discussion is based on this argument.

\subsection{Is the spur a part of the disk?}

The \HI\ velocity field around the optical galaxy shows the structure
expected for a disk with well-ordered rotation. The first question
that comes to mind is whether the spur is an extension of the \HI\
disk of \ugca. The outer \HI\ disk may not show well-ordered rotation
as a result of tidal interaction with IC~342, or because it has not
had the time to settle into a disk.  However, if the spur is to be
considered an extension of the disk, the orbits of \HI\ in the spur
should be reasonably circular around the optical galaxy and
approximately in the plane of the observed \HI\ disk. It should be
noted that some dwarf galaxies appear to have significant warps in
their outer \HI\ disks, suggested by their twisted \HI\ isophotes and
a change in P.A. of the velocity field.  Examples are DDO~168
\citep{Broeils1992,Stil2002a,Stil2002b}, UM~439 \citep{Vanzee98a}, and
NGC~4449 \citep{Bajaja94,hunter98}.

Three lines of argument indicate that the gas in the spur does not
satisfy these requirements. First, the spur partially overlaps in
position on the sky with the \HI\ disk, but not in velocity. This
implies that at least this part of the spur is significantly out of
the plane of the disk, because the orbits of gas in the spur cannot
cross the orbits of gas in the disk. The second argument follows from
the difference in the line of sight velocity between the spur and the
disk in the direction of the optical center of \ugca. Although the
rotation curve analysis of the velocity field did not uniquely define
the position of the center of mass of \ugca, it is reasonable to
assume that the center of mass is close to the optical center, as it
is for larger disk galaxies. The correspondence between the morphology
of the optical and the \HI\ isophotes supports this assumption. The
velocity difference $\Delta v$ is therefore the radial component of
the orbital velocity of gas in the spur, with respect to the dynamical
center of \ugca. For \ugca, $\Delta v \approx 40\ \kms$ is about
one third of the circular velocity of the outer disk at radius $8'$
(6 kpc). This large radial velocity component relative to the local
circular velocity implies that the orbit is significantly
non-circular. 

The first and second arguments combined imply that the part of the
spur that is observed in the direction of the optical center is
probably further from the dynamic center than the observed radius of
the \HI\ disk ($R_{\rm disk} = 6\ \rm kpc$). The distance of this part of
the spur from the plane of the disk is then at least $R_{\rm disk}
\cos(i) = 3.9\ \rm kpc$ for the adopted inclination $i = 50^\circ$.  

The third argument that the spur is not an extension of the \HI\ disk
is that differential rotation would wind-up the spur on an orbital
time scale $2 \pi R_{\rm disk}/v_{\rm c} \sim 3 \times 10^8\ \rm yr$. The linear
appearance of the spur would be highly coincidental because it would
exist only for a short time.

The conclusion is that the \HI\ spur is not an extension of the \HI\
disk.  This also excludes the possibility that the spur is an \HI\
spiral arm similar to those observed in the extended outer \HI\ disks
of some dwarf galaxies, such as NGC~2915 \citep{Meurer96} and DDO~47
\citep{Walter2001,Stil2002a}.  At least some \HI\ in the spur is found
to follow non-circular orbits at a high inclination with respect to
the plane of the \HI\ disk. \ugca\ is not unique in this
respect. Highly structured outer \HI\ with kinematics different from a
disk that coincides with the optical galaxy has been reported for
NGC~4449 \citep{hunter98,hunter99}, IZw~18 \citep{Vanzee98a}, IIZw~40
\citep{Vanzee98b} and NGC~1569 \citep{Stil1998,Stil2002c}.  IIZw~40 is
a clear example of a merger in progress. The \HI\ filaments in
NGC~4449 may be the result of tidal interaction.  However,
\citet{Vanzee98a} suggested that the extended \HI\ in IZw~18 is a
remnant of the nascent cloud from which the galaxy formed. There is no
known companion of NGC~1569 that could have caused the disturbed
kinematics in its outer regions through tidal interaction. These
examples show that there may not be a single origin of the disturbed
\HI\ kinematics in these systems. It is not possible in general to
decide unambiguously which process occurs from \HI\ observations of a
particular galaxy.

\subsection{The spur as a tidal tail}

\begin{figure}
\caption{ Mosaic of our two DRAO observations of the pair IC~342 and
\ugca\  showing the size and orientation of the spur in relation to
the size and mutual distance of this pair of galaxies. The \HI\ column
density is shown in grayscales, linearly from $2 \times 10^{20}\ \rm
cm^{-2}$ to $2 \times 10^{21}\ \rm cm^{-2}$. At a distance of 2.6 Mpc,
1 degree corresponds with 45 kpc. The faint extended emission between
\ugca\  and IC~342 is Galactic \HI.}
\label{UGCA86_mosaic}
\end{figure}

The proximity of IC~342 suggests that the spur may be the result of
tidal interaction with this much larger neighbor. If the spur is a
tidal tail induced by IC~342, some inferences can be made based on the
kinematics and morphology of the spur. First, the velocity of the spur
in the direction of the center of \ugca\  is blue shifted relative to
the \HI\ disk. A tidal tail is expected to have a radial velocity
component away from the center.  This implies that the blue shifted
spur gas should be located in front of the \HI\ disk. Second, the
3-dimensional space velocity of the gas in a tidal tail is expected to
vary smoothly with position along the tidal tail because adjacent test
particles in the tidal tail always experienced similar accelerations
during the interaction. This continuity of the spacial velocity should
be seen in the data as a smooth gradient of the line of sight velocity
along the tidal tail. The observed variation of the line of sight
velocity along the {\it minor} axis of the spur is not necessarily in
conflict with this expectation. The minor-axis velocity gradient may
be the result of a projection effect where different parts of the
tidal tail are seen close together on the plane of the sky. For this
to be the case, the tidal tail must be in a plane that is observed
nearly edge-on.  The orientation of the spur along the minor axis of
the optical and \HI\ isophotes of the disk of \ugca\  further means
that the plane of the tidal tail makes a large angle (close to
$90^\circ$) with the plane of the disk, no matter what the inclination
of the disk might be.  Third, the fact that the spur extends on the
sky in the direction of IC~342, and the necessity that the spur is in
a plane observed nearly edge-on, mean that IC~342 must be close to the
plane of the spur.  Fourth, if different parts of the tidal tail are
to be seen in projection on the sky along the entire spur, the tip of
the spur that appears most distant from \ugca\  is a point where the
line of sight is approximately a tangent to the tidal tail. If this is
the case, one would expect that the spur is brighter at this point and
that the velocity gradient along the minor axis vanishes. This
expectation is confirmed by Figure~\ref{gray:minorslice}.

The inference that IC~342 is in the same plane as the tidal tail is an
interesting argument in favor of an interpretation of the spur as a
tidal tail induced by IC~342. However, significant problems exist with
this interpretation. If the spur is a tidal tail in a nearly polar
plane, significant warping of the outer \HI\ disk of \ugca\ is also
expected to occur. This warp should manifest itself by a change in
P.A. of the \HI\ isophotes and the velocity field of the disk. No such
variation in P.A. is observed in Figure~\ref{gray:hicolumn} and
Figure~\ref{gray:velfield}.  \citet{Vanzee98a} similarly argued that
the component \HI-SX in IZw~18 is not a tidal tail because of lack of
continuity in the velocity field between the components \HI-A and
\HI-SX. The necessity of a tidal tail in a plane that is perpendicular
to the plane of the disk is by itself a problem because
low-inclination encounters seem to be most effective in the formation
of tidal tails \citep{Toomre1972}.  Finally, in the inferred geometry
the tidal tail appears in the direction of IC~342, with no evidence
for a counter tail.  This is contrary to simulations of tidally
interacting disk galaxies, where the counter tail is the most
prominent tidal feature \citep{Toomre1972,Barnes1988}. Detailed
modeling specific to this system is required to make these arguments
conclusive.

\subsection{The spur as a trigger of star formation}

\ugca\  currently experiences a period of enhanced star formation. The
detection of \ugca\  in the radio continuum, especially at 408 MHz,
indicate exceptional star forming activity for a dwarf galaxy. Further
evidence for significant recent star formation is seen in the numerous
\HII\ regions and the cometary knot.

It is tempting to associate the spur with the current star formation
in \ugca.  Gas falling into the disk as suggested by \citet{Ri91} and
\citet{Braun00} may have triggered the recent episode of star
formation.  Indirect evidence that this may have occurred is found in
the region with anomalous velocity shown in
Figure~\ref{holeslice-fig}. This region occurs on the side of \ugca\
that displays the most intense star formation. If the kinetic energy
of this region is supplied by stellar winds and supernovae, the
equivalent of more than 200 supernova explosions is required.  The
integrated star formation rate of \ugca\ of $0.014\ \rm M_\odot$
implies a supernova rate of $1 \times 10^{-4}$ supernovae per year for
the entire galaxy, assuming a Salpeter IMF with low mass cut-off
$0.01\ \rm M_\odot$ and high-mass cut-off $100\ \rm M_\odot$.  The
main \HII\ regions are found on the periphery of the anomalous
velocity region, not near the center. The power and the location of
the star formation regions suggest that the observable population of
high-mass stars is not the source of the anomalous velocity
region. The impact of high-velocity clouds in a galactic disk has been
suggested as an alternative energy source for super shells with very
high energy requirements \citep{tenorio1986,tenorio1987}. The presence
of the spur and a kinematically disturbed region in the disk that
requires more energy input than the observable stars can provide,
lends some credibility to the suggestion of gas falling into \ugca.

\section{Conclusions}

We present \HI, and 1.4 GHz and 408 MHz continuum observations of the
Magellanic dwarf irregular galaxy \ugca\ with the DRAO
interferometer. The \HI\ morphology of \ugca\ is characterized by a
rotating disk with inclination $\sim 50^\circ$, but a significant
misalignment exists between the optical/\HI\ major axis and the
kinematic major axis.  In addition to this, a narrow spur of \HI\
extends to the northwest, up to $12'$ (9 kpc) from the center of
\ugca. The spur is oriented along the apparent minor axis of the disk
of \ugca, and its kinematics are significantly different from the
disk.  The spur partially overlaps with the disk on the sky. The
velocity difference between the spur and the disk towards the center
of \ugca\ is $\Delta v \approx 40\ \kms$. In this direction we measure
directly the radial component of the orbital velocity of gas in the
spur. This large radial component, one third of the circular velocity
at the edge of the \HI\ disk, is evidence that the orbits of gas in
the spur are significantly non-circular. Also, the overlap of the spur
and the disk on the sky, and the modest inclination of the disk imply
that part of the spur is located well outside the plane of the disk,
but no evidence for a warp of the outer \HI\ disk is found.  A
well-defined velocity gradient is observed along both the major and
the minor axes of the spur.

The mean \HI\ velocity dispersion of the disk $\sigma_{\rm HI} = 8.8\
\kms$ is similar to what is observed in other (dwarf)
galaxies. However, the velocity dispersion of \HI\ in the spur shows a
large range, from $10\ \kms$ to $30\ \kms$.

\ugca\  is also detected in the continuum at 1.4 GHz and 408 MHz. The
1.4 GHz continuum image shows the three largest \HII\ region
complexes, and an extended component that covers the entire optical
disk of \ugca. At 408 MHz, the galaxy is unresolved. The global
spectral index between 408 MHz and 1.4 GHz is
$\alpha=-0.34^{+0.27}_{-0.20}$ ($S_\nu\propto\nu^\alpha$).

The interpretation of the spur as a tidal tail induced by IC~342,
following an earlier suggestion by \cite{Rots79} is considered. In
order to explain the observed \HI\ kinematics of the spur, the tidal
tail must be in a plane that is nearly perpendicular to the disk,
although no warping of the outer \HI\ disk is observed. Also, the spur
is oriented in the direction of IC~342, which means that the prominent
tidal tail is located in between the two interacting galaxies.
Detailed modeling of the interaction between IC~342 and \ugca\  is
necessary to address these objections against a tidal tail
interpretation of the spur.

If the spur is not a tidal tail, it is likely gas in an extended,
non-circular orbit around \ugca. This gas may then be a remnant of
the from the formation of \ugca\  that has not yet settled into the
disk, or an unlucky small dwarf galaxy that was tidally disrupted by
\ugca.

\section*{Acknowledgements}

The Dominion Radio Astrophysical Observatory is operated as a national
facility by the National Research Council of Canada.  This research
made use of the NASA/IPAC Extragalactic database (NED), which is
operated by the Jet Propulsion Laboratory at Caltech, under contract
with the National Aeronautics and Space Administration. JMS is
grateful to dr. A. R. Taylor for financial support during this
work. The authors thank dr. H. Lee for his comments on the manuscript.

\end{document}